\documentclass[aps,prl,showpacs,amssymb,twocolumn,floatfix]{revtex4}
\usepackage{epsfig}
\usepackage{times}
\begin{document}
\title{Localized transverse bursts in inclined layer convection}
\author{Karen E. Daniels$^1$, Richard J. Wiener$^2$, and Eberhard
Bodenschatz$^1$}
\affiliation{$^1$Laboratory of Atomic and Solid State Physics, 
Cornell University, Ithaca, NY 14853, $^2$Department of Physics,
Pacific University, Forest Grove, OR 97116}
\date{\today}

\begin{abstract}
We investigate a novel bursting state in inclined layer thermal
convection in which convection rolls exhibit intermittent, localized, 
transverse bursts. With increasing temperature difference, the bursts 
increase in duration and number while exhibiting a characteristic
wavenumber, magnitude, and size. 
We propose a mechanism which describes the duration of the
observed bursting intervals and compare our results to bursting
processes in other systems.  
\end{abstract}


\maketitle

Bursting phenomena are a common feature of nonequilibrium systems.  
Various bursting mechanisms have been
described by Knobloch and Moehlis \cite{Knobloch:2000:BMH} and
characterized in terms of their recurrence properties and dynamic
range, distinguishing between temporally localized bursts which are 
spatially global and those which are spatially local. 
Some examples of the former are the breakdown of spiral vortices
in Taylor-Couette flow \cite{Coughlin:1996:TBC,Coughlin:1999:TBC}, 
fluctuations in heat transport in binary fluid convection
\cite{Sullivan:1988:NTD,Moehlis:1998:FSB}, and shear flow turbulence  
\cite{Waleffe:1998:TDC,Grossmann:2000:OSF}. While a
dynamical systems approach has been fruitful in the global case,
spatially localized bursts remain less understood. 

\begin{figure}
\centerline{\epsfig{file=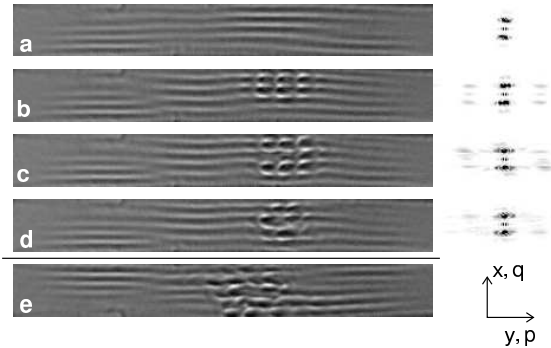, width=3in}}
\caption[Images and spectra of transverse bursts]{(a--d) Sequence of
background-divided shadowgraph images 
and associated contrast-enhanced power spectra 
for Cell 1 at $\theta = 78^\circ$ and $\epsilon =
0.08$. (a) quiescent, (b) modulations during 
early bursting, (c) developed bursting, and (d) end of burstlet. Times
correspond to points marked in Fig.~\protect\ref{B_f_pqmodes}.
Uphill direction is at left. 
(e) Disordered bursting at $\theta = 78^\circ$ and $\epsilon = 0.09$.  
Also see movies at \cite{EPAPS}. } 
\label{B_f_pics}
\end{figure}

In the Letter, we characterize a recently discovered
\cite{Daniels:2000:PFI} bursting phenomenon in inclined layer
convection (ILC) resulting from the presence of both buoyancy and
shear instabilities. Two features are intriguing: spatially
localized bursts of a characteristic size and the triggering of local
disorder within the burst without completely destroying the underlying
roll structure (see Fig.~\ref{B_f_pics}). Temporally, we observe
multiple cycles of transverse modulation, turbulence, and decay within
the bursts. Existing theory does not address either the
spatial or temporal dynamics. 

\begin{figure}
\centerline{\epsfig{file=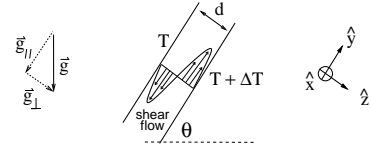, height=0.75in}}
\caption{Schematic diagram of inclined layer convection profile.}
\label{B_f_profile}
\end{figure}

Inclined layer convection is a variant of Rayleigh-B\'enard (thermal)
convection \cite{Bodenschatz:2000:RDR}, in which a fluid layer is
heated from one side and and cooled from the other.  Tilting this
layer by an angle $\theta$ (see Fig.~\ref{B_f_profile}) results in a
base state which is a 
superposition of a linear temperature gradient and a shear flow up
along the hot plate and down along the cold. When heated beyond a
critical 
temperature difference $\Delta T_c$, the fluid convects due to the
buoyancy of the hot fluid.
As $\theta$ is increased, the buoyancy provided by the perpendicular
component of gravity $g_\perp \equiv g \cos \theta$ becomes weaker and
the shear flow provided by $g_\parallel$ becomes stronger. 
Above a codimension-two point at $\theta_c$, the primary instability
is due to this shear flow instead of buoyancy
\cite{Hart:1971:SFD,Clever:1977:ILC,Fujimura:1993:MMC}. Interesting 
bursting behavior has been observed in the vicinity of this
codimension-two point \cite{Daniels:2000:PFI}.

We perform experiments in high pressure CO$_2$ in an apparatus
similar to that described in \cite{deBruyn:1996:ASR}, modified to
allow for inclination. The gas was at a
pressure of $(48.26 \pm 0.01)$ bar regulated to $\pm 0.005$ bar with a
mean temperature of $(25.00 \pm 0.05) ^\circ$C regulated to $\pm
0.3$mK. Our three convection cells were of height $d = 778 \pm 2 \mu$m
and length $97.3 d$. The widths in the $\hat{\bf x}$ direction are
$L_1 = 10.4d$ for Cell 1, $L_2 = 20.9 d$ for Cell 2, and $L_3 = 31.0
d$ for Cell 3. These parameters give a vertical viscous diffusion
time of $\tau_v = d^2/\nu = (4.50 \pm 0.02)$ sec, a Prandtl number
$\sigma = 1.301 \pm 0.001$, and weakly non-Boussinesq conditions ($Q =
0.2$ to 0.8, as described in \cite{Bodenschatz:2000:RDR} for
horizontal fluid layers). The planform of the convection pattern was
observed via the shadowgraph technique \cite{deBruyn:1996:ASR} using a
digital camera. The two control parameters are the angle $\theta$ and
the nondimensionalized temperature difference $\epsilon \equiv {\Delta
T \over \Delta T_c} -1$. Images were collected at 27 frames/$\tau_v$
with runs of duration at least 1000 $\tau_v$ at various values of
$\epsilon$, $\theta$.

\begin{figure}
\centerline{\epsfig{file=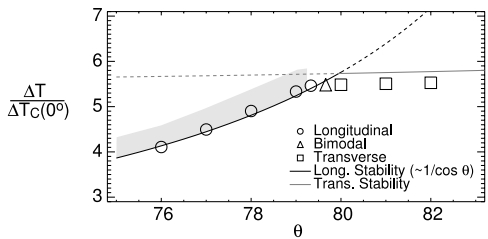, width=2.5in}}
\caption[Phase diagram of codimension-two point]{Phase diagram near
the codimension-two point of ILC, for 
Prandtl number $\sigma = 1.301$.  Stability curves are from
\cite{Pesch}; solid lines are primary instabilities and dashed
lines are extensions. Shaded area corresponds to data collection on
bursts.} 
\label{B_f_phaseplot}
\end{figure}

Fig.~\ref{B_f_phaseplot} shows the phase diagram for ILC in the
vicinity of the codimension-two point, with onset measured
experimentally (data points) and compared to predictions from linear
stability analysis \cite{Pesch}. The localized transverse
bursts 
appear intermittently within the underlying longitudinal rolls as a
secondary instability for $\theta < \theta_c$.
These bursts are triggered within a secondary instability in which
there is already local growth of regions of high-amplitude
longitudinal convection, as seen in the sequence of images in
Fig.~\ref{B_f_pics}; a movie of the corresponding images are available
online \cite{EPAPS}. Within these 
regions, transverse modulations repeatedly grow and decay, becoming
disordered or turbulent in the process. Eventually, this cycle of
bursting ends and the system returns to quiescent, weak longitudinal
rolls. The phenomenon 
becomes more pronounced at large $\epsilon$, so that eventually the
whole cell is bursting. Each localized patch does not
spread and no global bursting was observed. 
At lower inclination (further from the codimension-two point),
the bursting is indistinguishable from
the crawling rolls described in \cite{Daniels:2000:PFI}. 

\begin{figure}
\centerline{\epsfig{file=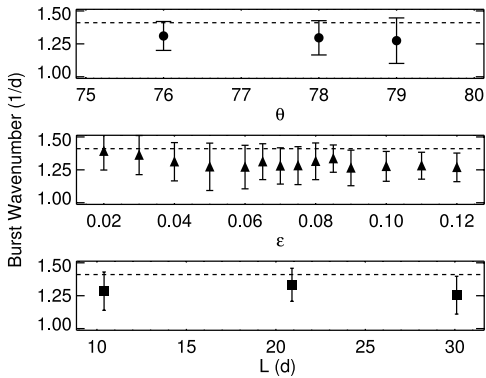,width=2.5in}}
\caption[Wavenumber of transverse bursts]{Wavenumber of transverse
bursts $p$ (determined from  
power spectra) as a function of $\theta$, $\epsilon$, and $L$. 
In each graph the data plotted are averaged over the
other two variables. Dashed lines are $p_c/2$ for $\theta =
80^\circ$.}
\label{B_f_burstk}
\end{figure} 

The measured wavenumber $p$ of the transverse modulations is shown in 
Fig.~\ref{B_f_burstk} and observed to be approximately constant. 
At $\theta = 80^\circ$, the transverse
rolls were determined to have a wavenumber of $p_c = (2.82 \pm
0.04)d$ at onset; the burst wavenumber $p$ is nearly subharmonic
with respect to this value. 
The appearance of this bursting phenomena close to the stability curve
for transverse rolls suggests that a related shear instability is
playing a role. Nonetheless, no mode resonant with $p_c$ is observed
as part of the bursting spectrum (see Fig.~\ref{B_f_pics}). 

\begin{figure}
\centerline{\epsfig{file=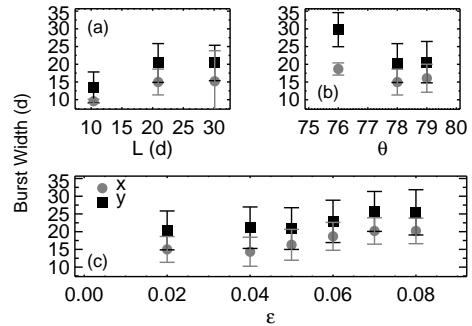,width=2.5in}}
\caption[Burst widths]{Burst width in $\hat{\bf x}$ (gray) and
$\hat{\bf y}$ (black) directions:
(a) vs. $L$ at onset of bursting for $\theta = 78^\circ$, (b)
vs. $\theta$ at onset of bursting in Cell 2, and (c) vs. $\epsilon$ in
Cell 2 at $\theta = 78^\circ$.} 
\label{B_f_burstxy}
\end{figure}

Using Fourier decomposition, we determined the area of each cell
occupied by transverse bursting and found the width of the region for
those instances in which only a single burst was present. As shown in
Fig.~\ref{B_f_burstxy}, the bursts generally
have a characteristic of size of $15d$ in the transverse
direction and $20d$ in the longitudinal. Bursts are smaller in Cell 1
(see Fig.~\ref{B_f_burstxy}a), 
where they fill the cell in the $\hat{\bf x}$ (transverse) direction,
and larger for $\theta = 76^\circ$ (Fig.~\ref{B_f_burstxy}b) and
increased $\epsilon$ (Fig.~\ref{B_f_burstxy}c)
Cell 2 and Cell 3 typically contain  
multiple bursts; therefore, we will utilize data from Cell 1 to focus
on the temporal behavior of single bursts. 
For Cells 1, 2, and 3 the onset of bursting was observed to occur at
$\epsilon = 0.075$, $\epsilon = 0.02$, and $\epsilon = 0.04$,
respectively \cite{onset}.

\begin{figure*}
\centerline{\epsfig{file=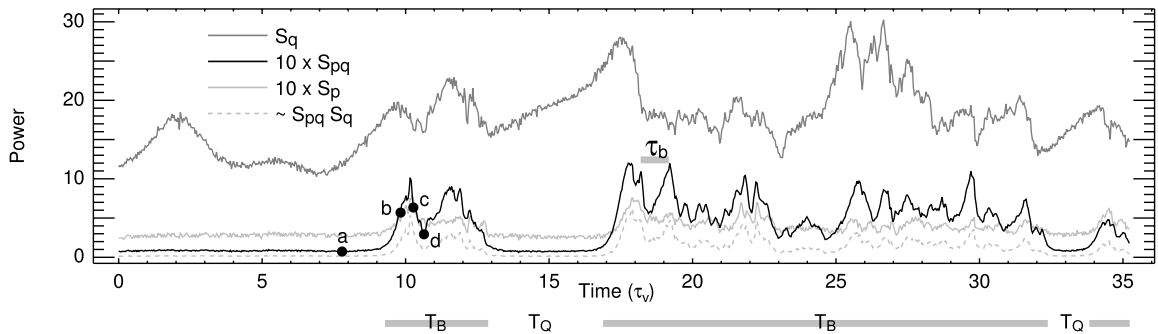, width=6in}}
\caption[Time trace of burst modes]{Sample segments of $S_q(t)$,
$S_{pq}(t)$, and $S_p(t)$ at 
$\epsilon = 0.08$ and $\theta = 78^\circ$ in Cell 1. Power is in
arbitrary units. Labeled black
dots correspond to the images  
in Fig.~\protect\ref{B_f_pics}. Gray bars indicated bursting; time
intervals $T_B$, $T_Q$, and  $\tau_b$ defined in text.} 
\label{B_f_pqmodes}
\end{figure*}

\begin{figure}
\centerline{\epsfig{file=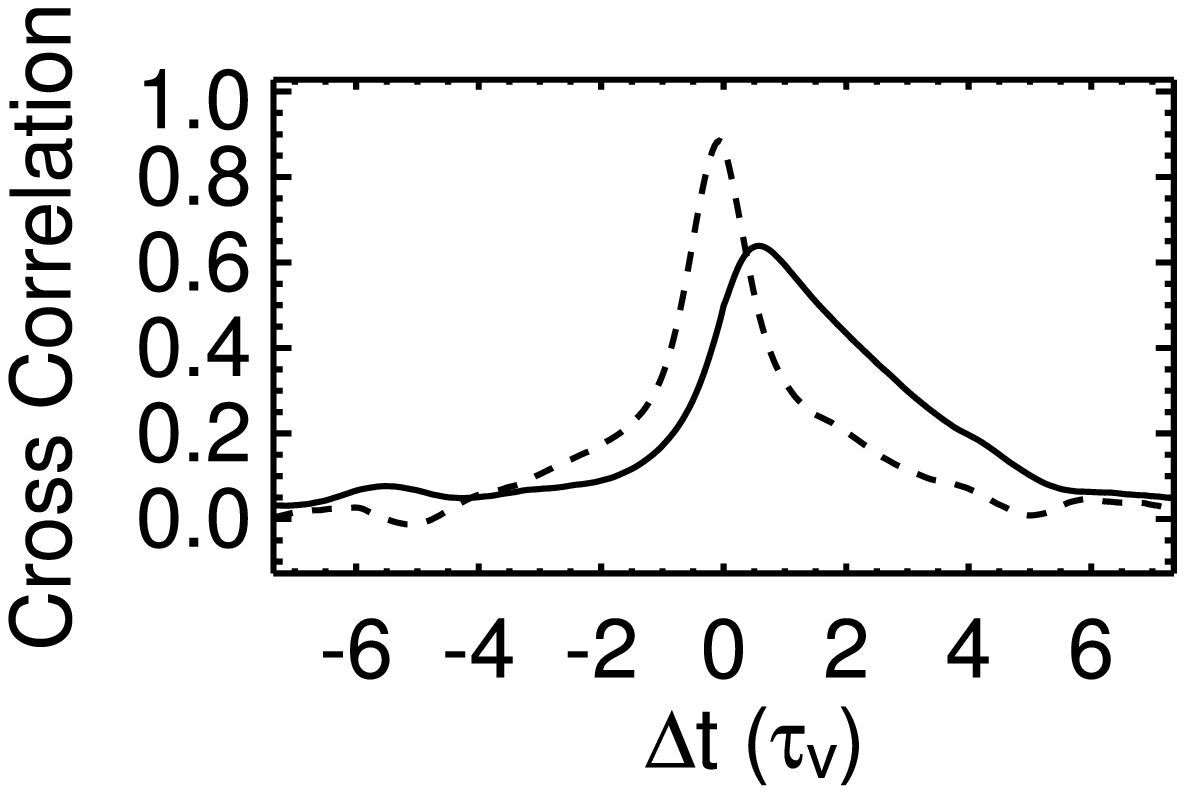, width=1.6in}
~~~\epsfig{file=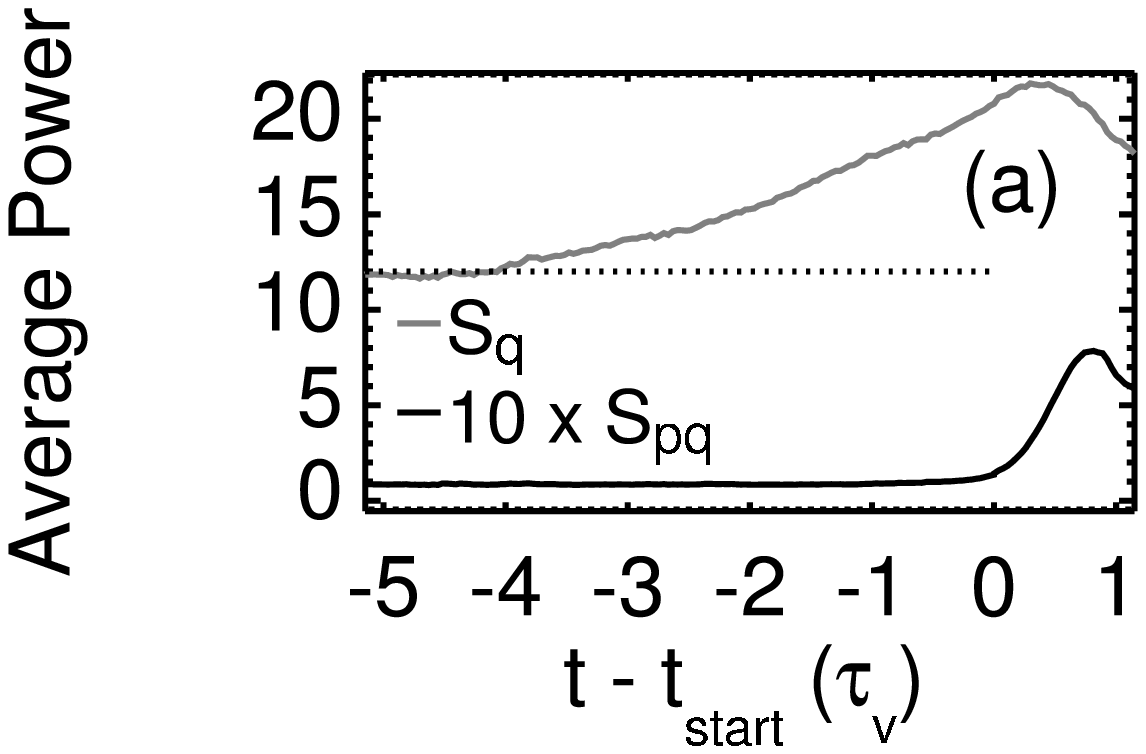, width=1.6in}}
\caption[Average burst power and mode cross-correlations]{
(a) Cross-correlation of ($S_{pq}$, $S_q$), dashed and ($S_{pq}$,
$S_p$), solid. (b) Ensemble average $\langle S(t - t_{\mathrm start})
\rangle$ over all bursts where $t_{\mathrm start}$ is the rising edge
of $S_{pq}$ for each burst. Both are plotted for Cell 1 at
$\epsilon=0.08$ and $\theta=78^\circ$. Power is in same
arbitrary units as Fig.~\protect\ref{B_f_pqmodes}.}
\label{B_f_aveburst}
\end{figure}

We describe the bursts based on the modes they excite in Fourier
space, using the power spectra of background-divided images 
as a function of time. Sample
spectra and their corresponding shadowgraph images 
are shown in Fig.~\ref{B_f_pics}, with three
prominent modes: a pure $\hat{\bf x}$ mode $q$, a pure 
$\hat{\bf y}$ mode $p$, and a mixed mode $pq$. 
The longitudinal rolls are composed of the pure $q$ mode, while the
transverse modulations are constructed from the $p$ and $pq$ modes.
Higher-order modes 
serve to create the characteristic shape of the bursts and
will be ignored here. A further simplification should be noted: the
shadowgraph technique integrates through the fluid layer ($\hat{\bf
z}$), while these 
bursting phenomena exhibit three dimensional behavior, particularly
when the modulations become disordered.

We define $S_q$, $S_{pq}$, and $S_p$ as the power in each of these
three peaks, taken as the total intensity in a 
fixed region of Fourier space. 
Fig.~\ref{B_f_pqmodes} shows time traces of the power in each
of the three modes. Prior to the beginning of the burst
$S_q$ typically increases in relative amplitude, as shown in
Fig.~\ref{B_f_aveburst}a. The beginning of a burst is characterized by the 
rapid growth of the transverse modulations (via $S_{pq}$). The $S_{p}$ 
mode is weaker and grows after 
a delay of about $0.5 \tau_v$ (see Fig.~\ref{B_f_aveburst}b), which
suggests that it is due to a 
nonlinear interaction or resonance condition 
between the $q$ and $pq$ modes. The simplest such nonlinearity would
be quadratic, and by multiplying $S_q(t)$ and $S_{pq}(t)$, we
obtain a good reproduction of the curve for $S_p(t)$, shown as the
gray dashed line in Fig.~\ref{B_f_pqmodes}. 

The time series in Fig.~\ref{B_f_pqmodes} show two temporal
features: (1) intermittent, alternating
periods of quiescence and bursting and (2) burstlets characterized by  
peaks and troughs within the bursting intervals. 
As $S_{pq}(t)$ rises at the beginning of a burstlet, the pattern
exhibits the increasingly
well-defined modulations shown in Fig.~\ref{B_f_pics}b. When 
$S_{pq}$ peaks, these modulations become spatially disordered on
short time scales, moving rapidly within the underlying rolls.
While the general roll pattern is retained in any single image (see 
Fig.~\ref{B_f_pics}c), the roll segments move turbulently within the
localized bursting region \cite{EPAPS}. At higher $\epsilon$ and
$\theta$, both of which increase the shear flow, this disorder is
increased (see Fig.~\ref{B_f_pics}d). 
From the disordered state, the modulations either decay ---
resulting in the end of the bursting interval --- or grow
again, creating another burstlet within the same interval. A
succession of such events is suggestive of the presence of a limit
cycle. Stochastic limit cycles have been previously described in
\cite{Knobloch:2000:BMH,Busse:1984:TTV}, relying on a random forcing
effect such as pressure fluctuations to produce random events with a
single, well-defined mean rate. 

\begin{figure}
\centerline{\epsfig{file=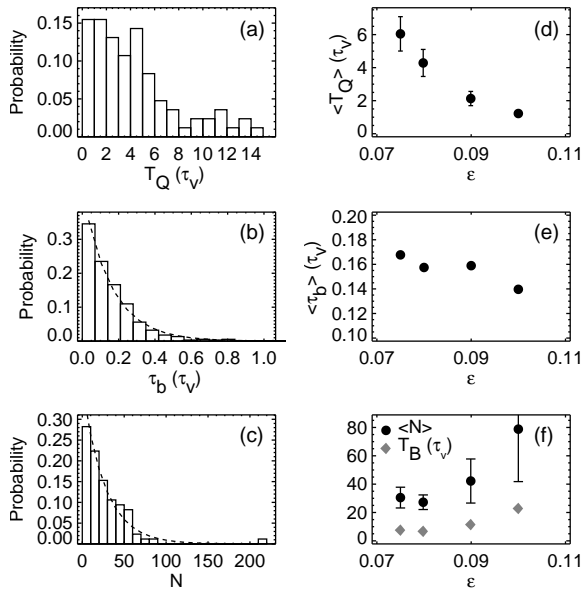, width=3in}}
\caption[Burstlet and wait time distributions]{(a,b,c) Sample PDFs of
$T_Q$, $\tau_b$, and $N$ respectively at $\epsilon = 0.08$ and
$\theta = 78^\circ$.  
(b) Dashed line is exponential distribution fit to data excluding the 
first bin. (c) Dashed line is Eqn. \protect\ref{e_ndist} 
plotted from the
measured mean. (d) Mean quiescent period as a function of
$\epsilon$ at $\theta = 78^\circ$. (e) Measured and
effective (from fit) mean burstlet period 
and (f) mean number of burstlets and mean bursting interval as a
function of $\epsilon$ at $\theta =  78^\circ$.} 
\label{B_f_bursttime}
\end{figure}

To determine the quiescent and bursting intervals, we set a threshold for
$S_{pq}(t)$ above which the system was considered to be
bursting. Because the onset and decay of the burst is sharp, the
results were not sensitive to the choice of threshold over a
reasonable range of values. Using this information, we examined the
duration of quiescent 
intervals $T_Q$, bursting intervals $T_B$. Burstlets are separated by
intervals $\tau_b$ within the $T_B$, with burstlet peaks identified by
local maximum.  Examples of the determined bursting 
intervals are shown by the gray bars at the bottom of
Fig.~\ref{B_f_pqmodes}. Sample 
probability distribution functions (PDFs) and mean values as 
a function of $\epsilon$ are plotted in Fig.~\ref{B_f_bursttime}. 
The quiescent intervals $T_Q$
were longest close to the onset of bursting, and decreased until they
were no longer detectable at higher $\epsilon$. Conversely, the
bursting intervals $T_B$ grow with $\epsilon$. The
combined effect is that at high $\epsilon$ the whole cell is in a
perpetual state of bursting since each localized burst lives longer
and new ones begin sooner. 
All of these trends hold at other values of $\epsilon$ and $\theta$ as
well. 

Such behavior distinguishes this bursting from the behavior of bursts
in Taylor-Couette flow \cite{Coughlin:1996:TBC,Coughlin:1999:TBC}.
There, the bursting is attributed to a secondary instability whose
growth above a threshold triggers turbulence throughout the fluid. 
Once the turbulence has
begun it destroys the underlying rolls which were providing energy 
and thus it dies away after
a well-defined period of time. It appears that the initial transverse
modulations in ILC 
are such a secondary instability, although they 
appear to arise due to a growing primary mode which triggers their
onset, as shown in Fig.~\ref{B_f_aveburst}. However, instead of fully
bursting, the burstlets  
cause only local disorder and fail to globally trigger
turbulence. Because the modulation and rolls are not
completely destroyed (see Fig.~\ref{B_f_pqmodes}), they are readily
able to grow 
back up again after they decay. This creates bursts comprised of
multiple burstlets. As a result, $T_B$ is not
constant as was observed for the Taylor-Couette bursts but instead 
increases with $\epsilon$, as shown in Fig.~\ref{B_f_bursttime}f.  
The mechanism for the growth of the Taylor-Couette bursts
\cite{Coughlin:1996:TBC,Coughlin:1999:TBC} provides for $\langle T_Q
\rangle \sim 1/\epsilon$ behavior based on a constant growth rate of
the secondary instability. It is possible that a similar mechanism is
at work since a similar trend of longer quiescent periods at lower
$\epsilon$ is observed (see Fig.~\ref{B_f_bursttime}d.)

To create cycles of burstlets within each bursting interval, a second
bursting process must be at work as well. At each $\epsilon$ and
$\theta$ the PDF of $\tau_b$ (see Fig.~\ref{B_f_bursttime}b) exhibits a
negative exponential distribution  
\begin{equation}
{\cal P}(\tau_b) = { e^{- \tau_b / \langle \tau_b \rangle} \over 
{\langle \tau_b \rangle} }
\label{e_tbexp}
\end{equation}
where $\langle \tau_b \rangle$ is the mean waiting period of a Poisson 
process generating the bursts. Because the lowest $\tau_b$ bin is
underreported due to the finite sampling of the time series,
$\langle \tau_b \rangle_{\mathrm eff}$ is found instead by fitting the
negative exponential to all data except for the first bin. 

This description can also explain why $T_B$ isn't constant. In a 
Poisson process, the events are memoryless: the wait time for the 
next event is independent of how long the system has already been
waiting. If a new burstlet were not generated before the decay to
quiescence, the modulations would die away and the burst interval
would be over. Assuming a constant, unknown decay time $\tau_d$, the
probability distribution for $N$ burstlets for which each
$\tau_b$ is no more than $\tau_d$ is calculated using the cumulative 
distribution of Eqn.~\ref{e_tbexp}, 
${\cal P}(\tau_b \le \tau_d) =  1 - e^{-\tau_b/\tau_d}$.
\begin{equation}
{\cal P}(N) = 
 {( 1 - {e}^{-\tau_d /  \langle \tau_b \rangle} )^N 
\over 
{e}^{ \tau_d /  \langle \tau_b \rangle} -1 }
\label{e_Pn}
\end{equation}
This distribution contains a single parameter 
$\langle N \rangle \equiv  {e}^{ \tau_d /  \langle \tau_b
\rangle}$, which can be measured directly from the experimental ${\cal
P}(N)$. Eqn.~\ref{e_Pn} then reduces to
\begin{equation}
{\cal P}(N) = { ( \langle N \rangle  -1)^{N-1} \over \langle N \rangle^N}
\label{e_ndist}
\end{equation}
where $\langle N \rangle$ is the measured mean number of burstlets per 
interval.  This formulation allows for comparison with the
observed data with no fit parameters. Fig.~\ref{B_f_bursttime}c shows
good agreement with this model, as do plots at other parameter
values. Therefore, our results are consistent with the burstlets being
Poisson-distributed events. 

Transverse bursts in inclined layer convection show many consistent
and unexplained properties across a range of $\epsilon$ and 
$\theta$: spatial localization, transverse modulations of
uniform wavenumber, persistence of the roll structure while bursting,
and repeated cycles of growth and decay once triggered. Aspects of both
spatiotemporal chaos and turbulence appear to be relevant and these
results represent a first step in understanding 
the structure and dynamics of this novel bursting phenomenon. In
particular, more work is needed to understand both the local growth of 
longitudinal rolls which triggers the bursting and the implications of
a resonance condition between the $p$, $q$, and $pq$ modes.

We wish to thank W. Pesch and J. Brink for sharing results from
stability analysis and full numerical simulations and J. P. Sethna for
fruitful discussions. We are grateful to NSF for support under
DMR-0072077 and the IGERT program in nonlinear systems, DGE-9870631.


\begin{thebibliography}{17}
\bibitem[{ons()}]{onset}
\bibinfo{note}{In addition to different onsets, the three cells were also
  observed to have somewhat different burst strengths, with Cell 2 exhibiting
  the strongest bursts. These effects may be related to a resonance between the
  width of the cells and $q$, to finite-size effects, or to temperature
  inhomogeneities.}
\end{thebibliography}
\end{document}